\documentclass[fleqn,twoside]{article}
\usepackage[headings]{espcrc2}

\readRCS
$Id: espcrc2.tex,v 1.2 2004/02/24 11:22:11 spepping Exp $
\usepackage{graphicx}
\usepackage[figuresright]{rotating}

 \def\BA{\begin{eqnarray}}  
 \def\BE{\begin{equation}}
 \def\BF{\begin{figure}[htb]}
 \def\BT{\begin{table}[htb]}
 \def\EA{\end{eqnarray}}
 \def\EE{\end{equation}}   
 \def\EF{\end{figure}}
 \def\ET{\end{table}}

 \def\la{\langle}
 \def\ra{\rangle}

 \def\lsim{\mathrel{\rlap{\lower4pt\hbox{\hskip1pt$\sim$}}
     \raise1pt\hbox{$<$}}}         %less than or approx. symbol
 \def\gsim{\mathrel{\rlap{\lower4pt\hbox{\hskip1pt$\sim$}}
     \raise1pt\hbox{$>$}}}         %greater than or approx. symbol
\newcommand{\AmS}{{\protect\the\textfont2
  A\kern-.1667em\lower.5ex\hbox{M}\kern-.125emS}}

\title{
Nuclear Suppression of Dileptons at Large-$x_F$
}
\author{J.~Cepila\address[FJFI]{
Czech Techn. University in Prague, FNSPE,
B\v rehov\' a 7,11519 Prague, Czech Republic
}
\thanks{
This work was supported by the Grant LC 07048 (Ministry of Education of 
the Czech Republic).
}
and
J.~Nemchik\addressmark\address{
Institute of Experimental Physics SAS, 
Watsonova 47,04001 Ko\v sice, Slovakia
}
\thanks{
This work was supported in part by the 
Slovak Funding Agency, Grant 2/7058/27 and by Grants VZ M\v SMT 
6840770039 and LC 07048 (Ministry of Education of the Czech Republic).
}
}

\runtitle{Nuclear Suppression of Dileptons at Large-$x_F$}
\runauthor{J. Cepila and J. Nemchik}

\begin{document}

%%%%%%%%%%%%%%%%%%%%%%%%%%%%%%%%%%%%
\begin{abstract}
We study a significant nuclear suppression of the relative production rates
(p(d)+A)/(p+d(p)) for the Drell-Yan process
at large Feynman $x_F$. Since this is the region of minimal values for 
the light-front momentum fraction variable $x_2$ in the target nucleus,
it is tempting to interpret this as a manifestation of coherence
or of a Color Glass Condensate.
We demonstrate, however, that this suppression mechanism is governed by the
energy conservation restrictions in multiple parton
rescatterings in nuclear matter. 
To eliminate nuclear shadowing effects coming from the coherence,
we calculate nuclear suppression in the light-cone dipole approach 
at large dilepton masses and at energy accessible at FNAL.
Our calculations are in a good agreement with data from the E772 experiment.
Using the same mechanism
we predict also nuclear suppression at forward rapidities in the RHIC
energy range. 
\vspace*{-0.30cm}
\end{abstract}
%%%%%%%%%%%%%%%%%%%%%%%%%%%%%%%%%%%%%%%

% typeset front matter (including abstract)
\maketitle

%*******************************************
\section{INTRODUCTION\label{intro}}
%*******************************************

Recent study of small-$x$ physics is realized 
at RHIC by measurements of high-$p_T$ particles
in $d+Au$ collisions at forward rapidities $y > 0$
\cite{brahms,star}.
If a particle with mass $m_h$ and transverse momentum
$p_T$ is produced in a hard reaction then the corresponding
values of Bjorken variable in the beam 
and the target are
$x_{1,2} = \sqrt{m_h^2 + p_T^2}\,e^{\pm y}/\sqrt{s}$.
Thus, at forward rapidities the target $x_2$ is $e^y$-
times smaller than at midrapidities. This allows
to study coherent phenomena (shadowing,
Color Glass Condensate (CGC)), which are expected 
to suppress particle yields.

However, a significant suppression at large $y$
for any reaction is observed so far
at any energy. Namely, all fixed target experiments
(see examples in \cite{knpsj-05}) have too low
energy for the onset of coherence effects since
$x_2$ is large. The rise of suppression with $y$
(with Feynman $x_F$) shows the same pattern as observed
at RHIC. This allows to favor
another mechanism common for all reactions arising 
at any energy. Such a common mechanism based on energy
conservation effects in initial state parton rescatterings
and leading to $x_F$
scaling of nuclear effects
was proposed in \cite{knpsj-05}.

The projectile hadron can be decomposed over different
Fock states. A nucleus has a higher resolution than a proton
due to multiple interactions and so
can resolve higher Fock components containing more constituents.
Corresponding
parton distributions fall off steeper at $x\to 1$ where
any hard reaction can be treated as a large
rapidity gap (LRG) process where no particle is produced
within rapidity interval $\Delta y = -\ln(1-x)$.   
The suppression factor as a survival probability for LRG
was estimated in \cite{knpsj-05}, $S(x)\sim 1-x$.
Each of multiple interactions of projectile partons 
produces an extra $S(x)$ and the weight factors are
given by the AGK cutting rules \cite{agk}.
As was shown in \cite{knpsj-05} the effective projectile
parton distribution correlates with the nuclear target
and reads
%
%
%%%%%%%%%%%%%%%%%%%%%%%%%%%%%%%%%%%%%%%%%%%%%%%%%%%%%%%%%%%%%%%%%%%%
\vspace*{-0.1cm}
 \BA
&&
f^{(A)}_{q/N}\bigl (x
%,q_T^2\bigr 
) =
C\,f_{q/N}\bigl (x
%,q_T^2\bigr 
)\, 
\nonumber\\
&&\times 
\frac{\int d~^2b\,
\left[e^{-x\,\sigma_{eff}T_A(b)}- e^{-\sigma_{eff}T_A(b)}\right]}
{(1-x)\int d~^2b\,\left[1- e^{-\sigma_{eff}T_A(b)}\right]}\, ,
\label{10}
 \EA
%%%%%%%%%%%%%%%%%%%%%%%%%%%%%%%%%%%%%%%%%%%%%%%%%%%%%%%%%%%%%%%%%%%%
%
%
where $T_A(b)$ is the nuclear thickness function,
$\sigma_{eff}$ was evaluated in \cite{knpsj-05} and
the normalization
factor $C$ is fixed by the Gottfried sum rule.

In this paper we study
the rise of suppression with $y$ ($x_1$) at FNAL reported 
by the E772 Collaboration \cite{e772} for the Drell-Yan (DY)
process. We predict similar nuclear effects also at RHIC
in the forward region expecting the same suppression pattern 
as seen at FNAL.

\vspace*{-0.2cm}

%*******************************************
\section{THE COLOR DIPOLE APPROACH\label{dipole}}
%*******************************************

The DY process in the target rest frame can
be treated as radiation of a heavy photon/dilepton
by a projectile quark.
The transverse momentum $p_T$ distribution of photon
bremsstrahlung in quark-nucleon interactions,
$\sigma^{qN}(\alpha,\vec{p}_T)$,
reads \cite{kst1}:
%
%===============================================================
\vspace*{-0.4cm}
 \BA
&&\hspace*{-.8cm}
\frac{d\sigma(qN\rightarrow \gamma^*\,X)}{d(ln\,\alpha)\,d^2p_T}
=
\frac{1}{(2\pi)^2}\,
\sum\limits_{in,f}\,
\int\,d^2 r_1\,d^2 r_2\,
\nonumber\\
&&\hspace*{-.8cm}
e^{i\vec{p}_T\cdot(\vec{r}_1 - \vec{r}_2)}
\Phi_{\gamma^* q}^*(\alpha,\vec{r}_1) 
\Phi_{\gamma^* q}(\alpha,\vec{r}_2)\,
\times\,
\label{20}
\\
&&\hspace*{-.8cm}
%\times\,
\frac{1}{2} \biggl \{\sigma_{\bar qq}(x,\alpha r_1)
+ \sigma_{\bar qq}(x,\alpha r_2)
- \sigma_{\bar qq}(x,\alpha |\vec{r}_1 - \vec{r}_2|)
\biggr \} ,
\nonumber
\vspace*{-0.2cm}
 \EA
%===============================================================
%
where $\alpha = p^+_{\gamma^*}/p^+_q$ and
the light-cone (LC) wave functions of the projectile
quark $q+\gamma^*$ fluctuation $\Phi_{\gamma^* q}^*(\alpha,\vec{r})$ are 
presented in \cite{kst1}.
Feynman variable is given as $x_F = x_1 - x_2$ and
in the target rest frame $x_1 = p^+_{\gamma^*}/p^+_p$.
For the dipole cross section $\sigma_{\bar qq}(x,\alpha r)$ in Eq.~(\ref{20}) 
we used parametrization from \cite{kmw-06}.

The hadron cross section is given convolving the 
parton cross section, Eq.~(\ref{20}) with
the corresponding parton distribution functions (PDFs)
$f_{q}$ and $f_{\bar{q}}$
\cite{kst1,krt-01},
%
%=============================================================
\vspace*{-0.2cm}
 \BA
&&\hspace*{-0.7cm}
\frac{d\sigma(pp\rightarrow \gamma^* X)}{dx_F\,d^2p_T\,dM^2}
=
\frac{\alpha_{em}}{3\,\pi\,M^2}
\frac{x_1}{x_1 + x_2}
\int_{x_1}^{1}
\frac{d\alpha}{\alpha^2}
\sum_q Z_q^2
\nonumber\\
&&\hspace*{-0.7cm}
\times
\biggl\{\hspace*{-0.07cm}
f_{q}\bigl (\frac{x_1}{\alpha},Q^2\bigr )
+ f_{\bar{q}}\bigl (\frac{x_1}{\alpha},Q^2\bigr )
\hspace*{-0.10cm}
\biggr\}
\frac{d\sigma(qN\to\gamma^*X)}{d(ln\,\alpha)\,d^2p_T}
\hspace*{-0.00cm} ,
\label{30}
\vspace*{-0.1cm}
 \EA
% =============================================================
%
where
$Z_q$ is the fractional quark charge,
PDFs $f_q$ and $f_{\bar q}$ are used
with the lowest order (LO) parametrization
from \cite{grv} at the scale 
$Q^2 = p_T^2 + (1 - x_1) M^2$
and the factor $\alpha_{em}/(3\pi\,M^2)$
accounts for decay of the photon into a dilepton.

\vspace*{-0.15cm}
%*******************************************
\section{DILEPTON PRODUCTION ON NUCLEAR TARGETS}
\label{dpA}
%*******************************************
%
%
%

The rest frame of the nucleus is very convenient for study
of coherence effects.
The dynamics of the DY process is regulated by the
coherence length $l_c$ related to the
longitudinal momentum transfer, $q_L = 1/l_c$,
which controls the interference between amplitudes of the
hard reaction occurring on different nucleons.
The condition for the onset of shadowing in a hard
reaction is sufficiently long coherence
length (LCL) in comparison with the nuclear radius, $l_c\gsim R_A$,
where
%
% ===================================================================
\vspace*{-0.3cm}
\BE
l_c 
= \frac{2E_q\,\alpha(1-\alpha)}
{(1 - \alpha)\,M^2 + \alpha^2\,m_q^2 + p_T^2}
%=
%\frac{1}{m_N\,x_2}
%\frac{(1 - \alpha)\,M^2}{(1 - \alpha)\,M^2 + m_q^2\,\alpha^2 + p_T^2}
\ ,
\label{40}
\EE
% ===================================================================
%
and $E_q = x_q s/2m_N$ and $m_q$ is the energy and mass
of the projectile quark. The fraction of the proton momentum
$x_q$ carried by the quark is related to
$x_1$ as $\alpha x_q = x_1$.
In the LCL limit the special advantage of the color
dipole approach allows to incorporate
nuclear shadowing effects via a simple eikonalization
of $\sigma_{\bar qq}(x,r)$
\cite{zkl},
i.e. replacing $\sigma_{\bar qq}(x,r)$
in Eq.~(\ref{20}) by $\sigma_{\bar qq}^A(x,r)$:
%
% ========================================================
\vspace*{-0.2cm}
\BE
\sigma_{\bar qq}^A =
2 \int d^2 b\,\biggl\{1 - \biggl [1 -
\frac{1}{2\,A}\,\sigma_{\bar qq}\,T_A(b)
\biggr ]^A\biggr\}\, .
\label{50}
\EE
% ========================================================
%
The corresponding predictions for nuclear broadening 
in DY reaction based on the theory \cite{kst1}
for LCL limit were presented in \cite{krtj-03}.

In the short coherence length (SCL) regime
the coherence length is shorter than the mean internucleon
spacing, $l_c\lsim 1\div 2\,$fm. 
In this limit there is no shadowing due to very short   
duration of the $\gamma^*+q$ fluctuation.
The corresponding theory for description of the   
quark transverse momentum broadening
can be found in \cite{jkt,jks-07}.
Here the probability distribution $W_A^q(\vec{k}_T,x_q,\vec{b},z)
= dn_q/d^2 k_T$ that a valence quark arriving
at the position $(\vec{b},z)$ in the nucleus $A$ will have
acquired transverse momentum $\vec{k}_T$ 
can be written in term of the quark density matrix,
$\Omega_{q}(\vec{r}_1,\vec{r}_2) = (b_0^2/\pi)\, 
\exp( - b_0^2 (r_1^2 + r_2^2)/2)$,
%
% ========================================================
\vspace*{-0.2cm}
\BA
&&\hspace*{-0.8cm}
W_A^q(\vec{k}_T,x_q,\vec{b},z)
= 
\hspace*{-0.1cm}
\frac{1}{(2 \pi)^2}
\int d^2 r_1 d^2 r_2\,e^{i\,\vec{k}_T\cdot
(\vec{r}_1 - \vec{r}_2)}
\nonumber \\
&&\hspace*{-0.4cm}
\times\,
\Omega_{q}(\vec{r}_1,\vec{r}_2)\,
%e^{- \frac{1}{2}\,b_0^2
%(r_1^2 + r_2^2)}\,
e^{- \frac{1}{2}\,\sigma_{\bar qq}
(x_q,\vec{r}_1 - \vec{r}_2)\,
T_A(\frac{\vec{r}_1 + \vec{r}_2}{2} + \vec{b},z)}
\, ,
\label{180}
\EA
% ========================================================
%
where $b_0^2 =
\frac{2}{3\,\la r_{ch}^2\ra}$
with $\la r_{ch}^2\ra = 0.79\pm 0.03\,$fm$^2$ representing the
mean-square charge radius of the proton.
$T_A(b,z)$ in Eq.~(\ref{180}) is
the partial nuclear thickness function,
$T_A(b,z) = \int_{-\infty}^{z}\,dz'\,\rho_A(b,z')$.

Transverse momentum acquired
by a quark on the nucleus, $W^{qA}(\vec{k}_T,x_q)$,
is obtained averaging 
Eq.~(\ref{180}) over the nuclear
density $\rho_A(b,z)$:
%  
% ========================================================
\vspace*{-0.2cm}
\BA
%&&\hspace*{-0.5cm}
%W^{qA}(\vec{k}_T,x_q) =
W^{qA} =
%\nonumber\\
%&&\hspace*{-0.5cm}
\frac{1}{A} \int d^2 b dz \rho_A(b,z)\,
W_A^q(\vec{k}_T,x_q,\vec{b},z)\, .
\label{210}
\EA
% ========================================================
%

\vspace*{-0.15cm}
The cross section,
$\sigma^{qA}(\alpha,p_T)$, for an incident quark to produce
a photon on a nucleus $A$ with transverse momentum
$p_T$ can be expressed convolving the probability function
$W^{qA}(\vec{k}_T,x_q)$ with the cross section
$\sigma^{qN}(\alpha,k_T)$ (see Eq.~(\ref{20})),
%
% ========================================================
%\vspace*{-0.1cm}
\BE
\sigma^{qA}(\alpha,p_T)
=   
\int d^2 k_T W^{qA}(\vec{k}_T,x_q) \sigma^{qN}
(\alpha,\vec{l}_T)\, ,
\label{240}
\EE
% ========================================================
%
\vspace*{-0.1cm}
where $\vec{l}_T = \vec{p}_T - \alpha \vec{k}_T$.
To obtain the transverse momentum distribution for an incident proton
one should integrate over $\alpha$ similarly as in Eq.~(\ref{30}):
%
% ========================================================
\vspace*{-0.4cm}
 \BA
&&\hspace*{-0.7cm}
\frac{d\sigma(pA\rightarrow \gamma^*\,X)}{d\,x_F\,d^2p_T\,dM^2}
=
\frac{\alpha_{em}}{3\,\pi\,M^2}
\frac{x_1}{x_1 + x_2}
\int_{x_1}^{1}
\frac{d\alpha}{\alpha^2}
\sum_q Z_q^2
\nonumber\\
&&\hspace*{-0.7cm}
\times\,
\biggl\{f_{q}\bigl (\frac{x_1}{\alpha},Q^2 \bigr )
+ f_{\bar{q}}\bigl (\frac{x_1}{\alpha},Q^2 \bigr )\biggr\}\,
\sigma^{qA}(\alpha,p_T)\, .
\label{250}
 \EA
% =============================================================
%

Nuclear effects in $p+A$ collisions are usually investigated
via the so called nuclear
modification factor, defined as
%
% ========================================================
\vspace*{-0.2cm}
\BE
R_A(p_T,x_F,M) =
\frac{
\frac{d\sigma(pA\rightarrow \gamma^*\,X)}{d\,x_F\,d^2p_T\,dM^2}}
{A\,\,
\frac{d\sigma(pN\rightarrow \gamma^*\,X)}{d\,x_F\,d^2p_T\,dM^2}}\, ,
\label{270}
\EE
% ========================================================
%
where the numerator
is calculated in SCL and LCL regimes as described above.
Corrections for the finite coherence length was realized
by linear interpolation using nuclear longitudinal formfactor
\cite{knst-02} (for more sophisticate
Green function method see \cite{kst1,n-03}).

Note that at RHIC energy and at forward rapidities (large $x_F$)
the eikonal formula for LCL regime, Eqs.~(\ref{30}) and (\ref{50}),
is not exact since higher Fock components containing gluons
lead to additional corrections, called gluon shadowing (GS).
The corresponding suppression factor $R_G$ was derived
in \cite{knst-02,krtj-03} and included in calculations
replacing in Eq.~(\ref{50}) $\sigma_{\bar qq}$
by $R_G\,\sigma_{\bar qq}$. GS leads to reduction of the
Cronin effect \cite{knst-01} at moderate $p_T$ and to
additional suppression (see Fig.~\ref{fig5}).
 
For elimination of the coherence
effects one can study production of dileptons at large
$M$ (see Eq.~(\ref{40})) as has been realized
by the E772 Collaboration \cite{e772}.
Another possibility is to study the DY process at large
$x_1\to 1$, when also $\alpha\to 1$, and $l_c\to 0$ 
in this limit (see Eq.~(\ref{40})).

%
%******************************************************
\section{NUCLEAR SUPPRESSION AT FNAL ENERGIES}
%******************************************************
%

\begin{figure}[tbh]

\vspace*{-0.80cm} 
\includegraphics[scale=0.41]{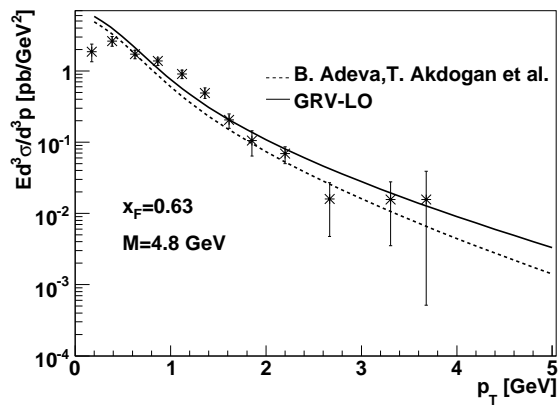}
\begin{center}

\vspace*{-1.3cm} 
\caption
{
Differential cross section of dileptons in $p+p$ collisions
at $x_F = 0.63$ and $M = 4.8\,$GeV vs. E886 data \cite{e886-pp}.
}
\label{fig1}
\end{center}
\vspace*{-1.3cm} 

\end{figure}

We start with
the DY process in $p+p$ collisions. 
Besides calculations based on Eq.~(\ref{30}) using
GRV PDFs \cite{grv} (see the dashed line
in Fig.~\ref{fig1}) we present by the solid line
also predictions using proton structure functions
from \cite{smc}.
Fig.~\ref{fig1} shows
a reasonable agreement of the model
with data from the E886 Collaboration \cite{e886-pp}.
This encourages us to apply the color dipole 
approach to nuclear targets as well.
%===============================================================
%\obr{fig1}{
%Differential cross section of dileptons in $p+p$ collisions
%at $x_F = 0.63$ and $M = 4.8\,$GeV vs. E886 data \cite{e886-pp}.
%}{8.0cm}{
%fig1}
%===============================================================

%===============================================================
%
%===============================================================
%\obr{fig2}{
%Ratio $R^{DY}(W/D)$ of Drell-Yan cross sections 
%on W and D vs. E772 data using long coherence length (LCL) and 
%short coherence length (SCL) limit and their interpolation (REAL) 
%for $6 < M < 7\,$GeV. The lower and upper series of curves
%are calculated with and without effects of the energy conservation,
%respectively.
%}
%{7.5cm}{fig2}
%===============================================================
 \begin{figure}[tbh]

\vspace*{-0.7cm} 
\includegraphics[height=6.7cm,width=8cm]{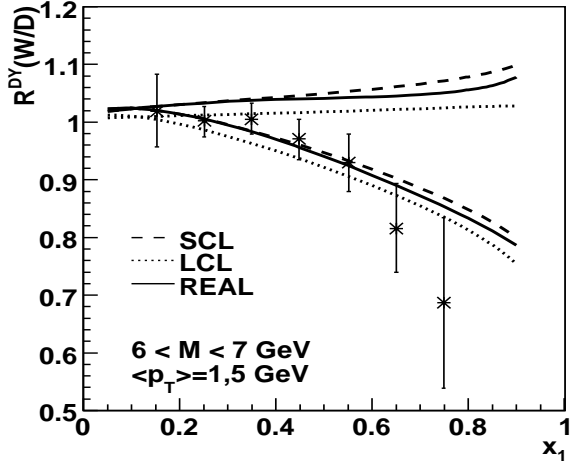}
\begin{center}

\vspace*{-1.3cm} 
\caption
{
Ratio $R^{DY}(W/D)$ of Drell-Yan cross sections 
on W and D vs. E772 data 
for $6 < M < 7\,$GeV.
Predictions correspond to the long (LCL) and 
short coherence length (SCL) regimes, and their interpolation (REAL). 
The lower and upper series of curves
are calculated with and without energy conservation effects,
respectively.
}
\label{fig2}
\end{center}
\vspace*{-1.5cm} 

\end{figure}
%===============================================================
%

The E772 Collaboration \cite{e772} found a significant suppression of
DY pairs at large $x_1$ (see Fig.~\ref{fig2}).
Large invariant masses of the photon allows to minimize
shadowing effects (see a small differences between lines
calculated in SCL and LCL regimes). If effects of energy
conservation are not included one can not describe a strong
suppression at large $x_1$. 
In the opposite case a reasonable agreement
of our model with data is achieved.
%===============================================================
%\obr{fig3}{
%The same as fig.~\ref{fig2} but for $7 < M < 8\,$GeV.
%}
%{7.5cm}{fig3}
%===============================================================
%\begin{figure}[tbh]
%
%\vspace*{-0.60cm} 
%\includegraphics[height=6.8cm,width=8cm]{fig3.eps}
%\begin{center}
%
%\vspace*{-1.3cm} 
%\caption
%{
%The same as fig.~\ref{fig2} but for $7 < M < 8\,$GeV.
%}
%\label{fig3}
%\end{center}
%\vspace*{-1.6cm} 
%
%\end{figure}
%===============================================================

\begin{figure}[tbh]

\vspace*{-0.65cm} 
\includegraphics[height=6.7cm,width=8cm]{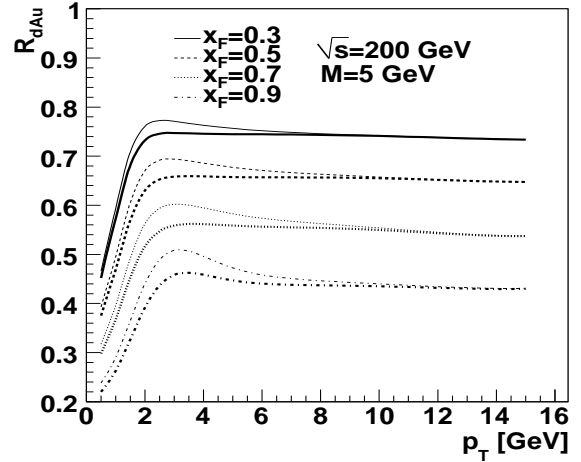}
\begin{center}

\vspace*{-1.35cm} 
\caption
{
Predictions for the ratio $R_{d+Au}(p_T)$ at $\sqrt{s} = 200\,$GeV
for several fixed values of $x_F$ with the energy 
conservation effects (thin lines). Thick lines
additionally include gluon shadowing effects.
}
\label{fig5}
\end{center}
\vspace*{-1.4cm} 

\end{figure}

Finally, we present also predictions for $p_T$ dependence
of the nuclear modification factor $R_{d+Au}$ at RHIC energy
and at several fixed values of $x_F$.
Similarly as in \cite{knpsj-05} instead of usual Cronin enhancement, a 
suppression is
found (see Fig. ~\ref{fig5}). 
The onset of isotopic effects at large $p_T$ gives a value 
$R_{d+Au}\sim 0.73\div 0.79$ and can not explain strong
nuclear effects.
The predicted
huge rise of suppression
with $x_F$ in Fig.~\ref{fig5} reflects much smaller survival probability
$S(x_F)$ at larger $x_F$ and can be tested in the future
by the new data from RHIC.
Note that effects of GS depicted in Fig.~\ref{fig5} 
by the thick lines lead to additional suppression
which rises with $x_F$.
%===============================================================
%\obr{fig4}{
%Predictions for the ratio $R_{d+Au}(p_T)$ at $\sqrt{s} = 200\,$GeV
%for several fixed values of $x_F$ when the effects of energy conservation
%are not included.
%}{7.5cm}{fig4}
%===============================================================
%\begin{figure}[tbh]
%
%\vspace*{-1.00cm} 
%\includegraphics[height=6.8cm,width=8cm]{fig4.eps}
%\begin{center}
%
%\vspace*{-1.3cm} 
%\caption
%{
%Predictions for the ratio $R_{d+Au}(p_T)$ at $\sqrt{s} = 200\,$GeV
%for several fixed values of $x_F$ without inclusion of the energy 
%conservation effects.
%}
%\label{fig4}
%\end{center}
%\vspace*{-1.6cm} 
%
%\end{figure}
%===============================================================
%===============================================================
%\obr{fig5}{
%The same as fig.~\ref{fig4} but with the effects
%of energy conservation (thin lines). Thick lines
%additionally include gluon shadowing effects.
%}{7.5cm}{fig5}
%===============================================================
%===============================================================
\vspace*{-0.3cm}

\section{SUMMARY}
\vspace*{-0.1cm}

We present unified approach to large $x_1$ ($x_F$) nuclear suppression 
based on energy conservation effects in multiple parton
rescatterings. 
We apply this approach for the DY process 
and explain well a significant suppression at large $x_1$
in accordance with the E772 data. The FNAL energy range and
large invariant masses of the photon allow to minimize
the coherence effects, what does not leave much room
for other mechanisms, such as CGC.
We predict a significant suppression also in $d+Au$ collisions
at RHIC in the forward region (see Fig.~\ref{fig5}).
At moderate $p_T$ we show an 
importance of GS effects and their rise with $x_F$.

\end{document}